**Observation of the spin splitting torque in a collinear antiferromagnet RuO$_2$**

H. Bai,[1,*] L. Han,[1,*] X. Y. Feng,[2,*] Y. J. Zhou,[1] R. X. Su,[1] Q. Wang,[1] L. Y. Liao,[1] W. X. Zhu,[1] X. Z. Chen,[1] F. Pan,[1] X. L. Fan[2,†] and C. Song[1,‡]

[1]Key Laboratory of Advanced Materials, School of Materials Science and Engineering, Tsinghua University, Beijing 100084, China

[2]The Key Lab for Magnetism and Magnetic Materials of Ministry of Education, Lanzhou University, Lanzhou 730000, China

Current-induced spin torques provide efficient data writing approaches for magnetic memories. Recently, the spin splitting torque (SST) was theoretically predicted, which combines advantages of conventional spin transfer torque (STT) and spin-orbit torque (SOT) as well as enables controllable spin polarization. Here we provide the experimental evidence of SST in collinear antiferromagnet RuO$_2$ films. The spin current direction is found to be correlated to the crystal orientation of RuO$_2$ and the spin polarization direction is dependent on (parallel to) the Néel vector. These features are quite characteristic for the predicted SST. Our finding not only presents a new member for the spin torques besides traditional STT and SOT, but also proposes a promising spin source RuO$_2$ for spintronics.

Current-induced spin torques not only enrich fundamental physics, but also provide efficient data writing approach for magnetic memories. The discovery of spin transfer torque (STT) brings about electrical switching of ferromagnetism, giving rise to the non-volatile magnetic random-access memory (STT-MRAM) with high speed and low consumption [1−5]. The longitudinal spin polarized current for STT is odd under time reversal ($\mathcal{T}$) and has high spin torque efficiency, due to the strong nonrelativistic ferromagnetic exchange splitting [1, 2, 4]. In contrast, transversal spin current with $\mathcal{T}$-even can be generated via the relativistic spin Hall effect (SHE) or/and the Rashba effect [6−8], which decouples reading and writing paths in MRAM and improves the device endurance. The resultant spin-orbit torque (SOT) has been extensively studied in the last decade for SOT-MRAM [9, 10].

Recently, a distinct spin splitting torque (SST) with the origin of nonrelativistic anisotropic spin band splitting was theoretically predicted in antiferromagnets [11−17]. In this scenario, transversal spin current with high spin torque efficiency is generated by the magnetic exchange splitting ($\mathcal{T}$-odd) and is independent on the spin-orbit coupling (SOC), which provides a unique opportunity to combine the advantages of STT and SOT. Meanwhile, SST shows the advantage of controllable spin polarization, which would expand the horizon of spin torque switching. For example, SST offers a different approach for generating out-of-plane spin polarization, and the efficiency is expected to be higher than other mechanisms relying on low crystal symmetry and magnetic ordering [18−21]. The experiments below provide evidence of the anisotropic spin splitting effect (ASSE) induced SST in a collinear antiferromagnet $RuO_2$, where the spin current direction is correlated to the crystal orientation and the spin polarization direction is dependent on the magnetic orientation, *i.e*., Néel vector of $RuO_2$.

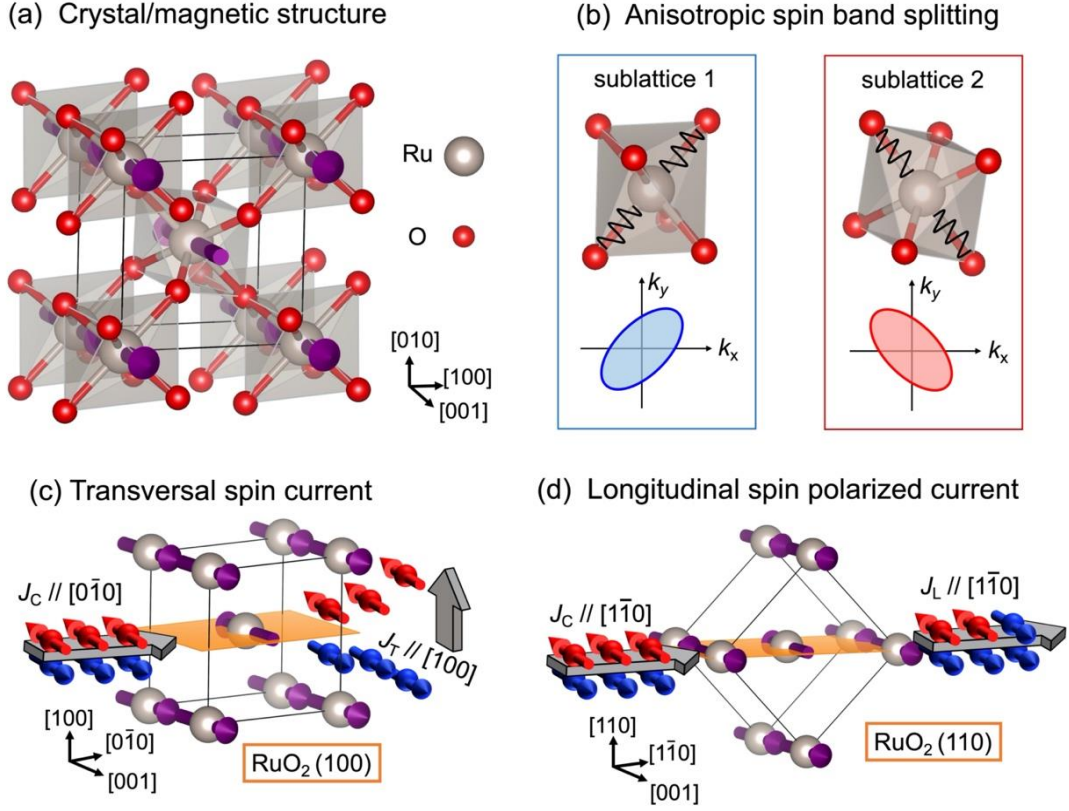

FIG. 1. (a) Crystal and magnetic structure of the rutile $RuO_2$. Grey and red spheres represent Ru and O atoms, respectively. Purple arrows mark the Ru local moments. (b) Schematic of the anisotropic spin band splitting in $RuO_2$. Oxygen octahedrons of two sublattices are rotated by 90º. The Ru atoms of two magnetic sublattices feel anisotropic octahedral crystal field (wavy black lines), leading to the anisotropic spin band splitting, as displayed by blue and red ellipses. (c) For the (100)-oriented $RuO_2$ film, transversal spin current ($J_T$) flowing along [100]-axis (out-of-plane) can be induced by the charge current along [0$\bar{1}$0]-axis. The spin polarization is parallel to the Néel vector ([001]-axis). (d) For the (110)-oriented $RuO_2$ film, spin polarized current ($J_L$) flowing along longitudinal direction can be generated by the charge current along [1$\bar{1}$0]-axis. The spin polarization is also parallel to the Néel vector.

RuO$_2$ was commonly considered as a paramagnet, until the recent finding of itinerant antiferromagnetism, with the Néel temperature above 300 K and the Néel vector aligned along [001]-axis [22, 23]. The crystal and magnetic structure of RuO$_2$ is shown in Fig. 1(a). RuO$_2$ is a rutile oxide with the P4$_2$/*mnm* space group, where Ru atoms occupy the center of stretched oxygen octahedrons [24]. Consequently, Ru atoms suffer from the octahedral crystal field, giving rise to anisotropic electronic structure and elliptical Fermi circles at $k_z$=0 [12], as illustrated in Fig. 1(b). Note that Ru atoms of opposite magnetic sublattices are surrounded by different directional oxygen octahedrons (with 90º rotation). Such a sublattice rotation in real space results in anisotropic spin band splitting in momentum space [Fig. 1(b)], ensuring RuO$_2$ an efficient spin splitter to generate spin current. Two typical configurations to generate spin current were proposed in Ref. [12]. For the (100)-oriented RuO$_2$ film, charge current applied along [0$\bar{1}$0]-axis can induce a transversal $\mathcal{T}$-odd spin current flowing along [100] direction (out-of-plane), and the spin polarization direction is parallel to the Néel vector ([001]-axis), as illustrated in Fig. 1(c). When the RuO$_2$(100) film is adjacent to a ferromagnetic layer, the transversal spin current will induce the spin splitting torque. Differently, for the (110)-oriented RuO$_2$ film, charge current applied along [1$\bar{1}$0]-axis produces only longitudinal spin polarized current, the transversal spin current is forbidden due to the symmetry, as displayed in Fig. 1(d) [25]. As a result, SST is absent for this case.

To probe SST experimentally, we deposited 12 nm thick (100)- and (110)-oriented RuO$_2$ films onto (100)-oriented yttria-stabilized zirconia (YSZ) and (100)-oriented MgO substrates, respectively. Then 8 nm thick ferromagnetic permalloy (Py) layer and 2 nm thick Al capping layer were in-situ deposited on the RuO$_2$ films. After fabricating into the device, we performed spin torque-ferromagnetic resonance (ST-FMR) measurements [25], which is a standard technique to detect spin current and spin torque. Note that ST-FMR is widely used to calibrate the spin current strength and distinguish the spin polarization direction [18, 20, 21, 26], which makes it pretty suitable for detecting SST. Compared to other detection techniques of charge-to-spin conversion, such as the harmonic Hall measurement, ST-FMR has the advantage of simplicity and large signal intensity [27]. In this work, two key parameters (spin torque efficiency $\theta_{SH}^{eff}$ and spin torque conductivity $\sigma_{SH}^{eff}$) are extracted from the ST-FMR signals to calibrate the charge-to-spin conversion efficiency. Herein, $\theta_{SH}^{eff}$ refers to the spin current density per unit current density, and $\sigma_{SH}^{eff}$ represents the angular momentum absorbed by the magnet per second per unit interface area per applied electric field. The detailed analysis methods of the ST-FMR signals are presented in the supplemental material [25].

Then we turn to the experimental results. We show in Fig. 2(c) and 2(d) the ST-FMR spectra of $RuO_2$/Py samples measured at $\varphi=45°$, which can be decomposed into symmetric ($V_S$) and antisymmetric components ($V_A$). The angle between the charge current and the external magnetic field is termed as $\varphi$. Note that the RF current $I_{RF}$ was estimated by calibrating the reflection coefficient ($S_{11}$) of the ST-FMR device, by which we conclude that $V_A$ of the ST-FMR data mainly arises from the Oersted field [25]. As a result, the ratio of $V_S/V_A$ is proportional to the spin torque efficiency [26]. Fig. 2(c) displays the ST-FMR spectrum of $RuO_2(100)$/Py sample. From the first glance, the amplitude of $V_S$ is comparable to $V_A$ ($V_S/V_A$=0.36), indicating the Py layer absorbs a strong spin torque [26]. The situation turns out to be dramatically different for the $RuO_2(110)$/Py, where the amplitude of $V_S$ is much smaller than $V_A$ ($V_S/V_A$=0.10), implying weak spin torque exerting on the Py layer, as displayed in Fig. 2(d). The angle-dependent ST-FMR measurements also draw the same conclusion [25]. We mention that this distinction shows the characteristic of SST, as shown in Fig. 2(a) and 2(b). For the RuO$_2$(100)/Py sample, both SST and SOT exist, giving rise to strong resonance and large $V_S/V_A$ ratio [Fig. 2(a)]. While for the RuO$_2$(110)/Py sample, the SST is absent, only weak resonance and small $V_S/V_A$ ratio are induced by SOT [Fig. 2(b)].

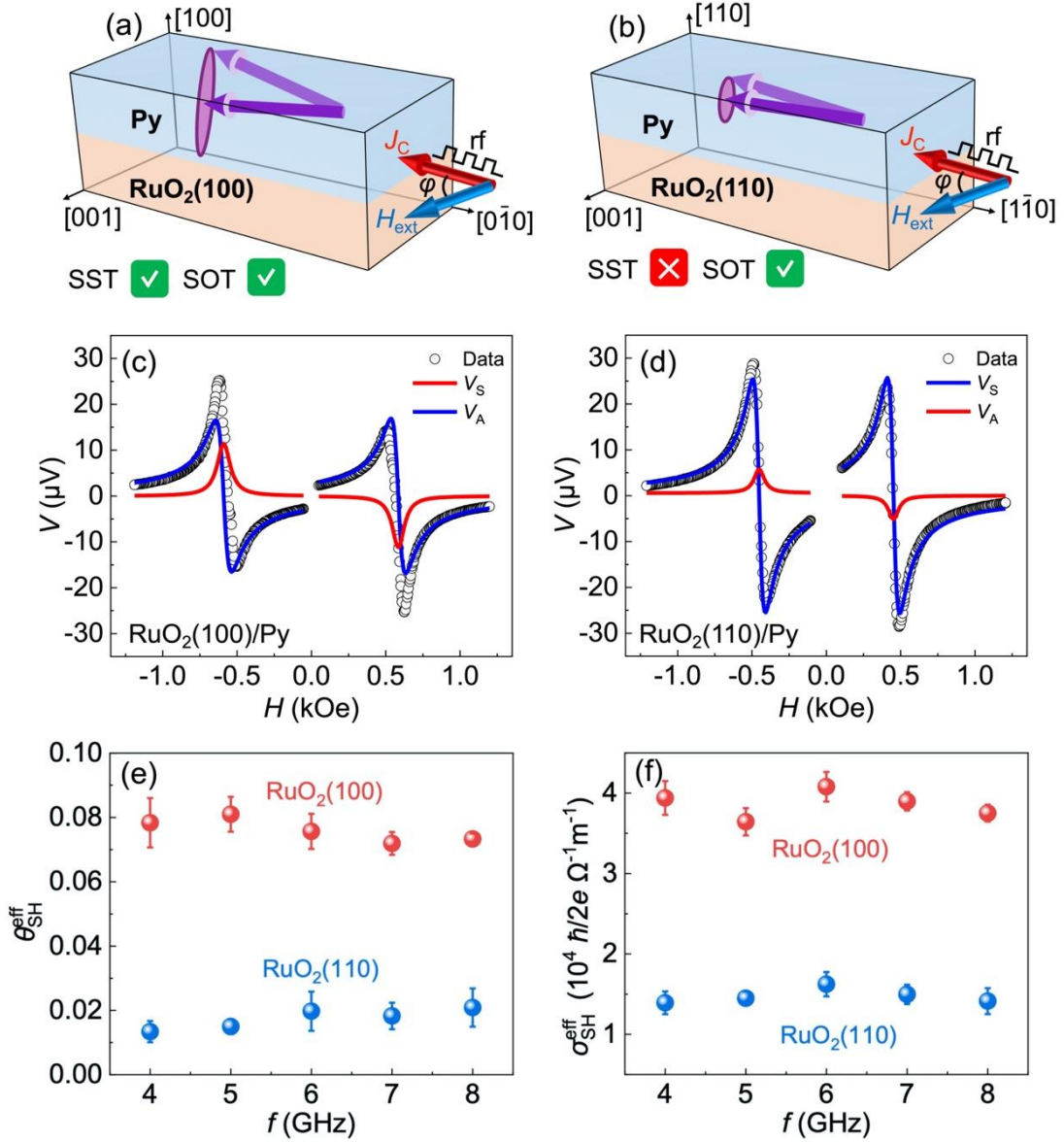

FIG. 2. (a, b) Schematic of ST-FMR measurements for the (a) RuO$_2$(100)/Py and the (b) RuO$_2$(110)/Py samples. For the RuO$_2$(100) film, both spin splitting torque (SST) and spin-orbit torque (SOT) exist, giving rise to a strong spin torque. For the RuO$_2$(110) film, SST is absent, only SOT contributes to a weak spin torque. The angle between the charge current ($J_C$) and the external magnetic field ($H_{ext}$) is termed as $\varphi$. (c, d) ST-FMR spectra of the (c) RuO$_2$(100)/Py and the (d) RuO$_2$(110)/Py samples measured at $\varphi=45°$ and $f=6$ GHz. The input power $P_{in} = 15$dBm. Circles are the raw data; red and blue lines represent the symmetric ($V_S$) and the antisymmetric ($V_A$) components, respectively. (e, f) The calculated (e) spin torque efficiency $\theta_{SH}^{eff}$ and the (f) spin torque conductivity $\sigma_{SH}^{eff}$ at different microwave frequencies in RuO$_2$(100) and RuO$_2$(110) films. The error bars are from three repetitive measurements.

In order to quantitatively compare the spin torque efficiency of the two samples, $\theta_{SH}^{eff}$ is calculated by Eq. (S2). Corresponding data are shown in Fig. 2(e). For both (100)- and (110)-oriented RuO$_2$ films, $\theta_{SH}^{eff}$ exhibit little variation as a function of microwave frequency, excluding artifacts rooted from the specified microwave frequency. For the RuO$_2$(100) film $\theta_{SH}^{eff}$ is about 0.08, which is much larger than that of RuO$_2$(110) film ($\theta_{SH}^{eff} \sim 0.02$) and is comparable to that of the typical heavy metal Pt [26]. Similar crystal orientation-dependent $\theta_{SH}^{eff}$ was observed in rutile paramagnetic IrO$_2$ [28]. Furthermore, the effective spin torque conductivity ($\sigma_{SH}^{eff}$) is calculated for the two samples by estimating the RF current [25]. Corresponding data are plotted in Fig. 2(f). For the (100)-oriented RuO$_2$ film, $\sigma_{SH}^{eff}$ is up to $4 \times 10^4$ $\hbar$/2e $\Omega^{-1}$ m$^{-1}$, which is around three times as large as that of the (110)-oriented RuO$_2$ film. Note that we calibrated the interfacial spin transparencies of these two samples by the ferromagnetic resonance measurement, the effective spin mixing conductance $g_{eff}^{\uparrow\downarrow}$ of them are comparable [25], thereby the enhanced spin torque efficiency in RuO$_2$(100) sample can be ascribed to an extra spin current generation mechanism in addition to SHE. The generation of SST due to the anisotropic spin band splitting is a reasonable explanation, which is also in agreement with the theoretical prediction [12].

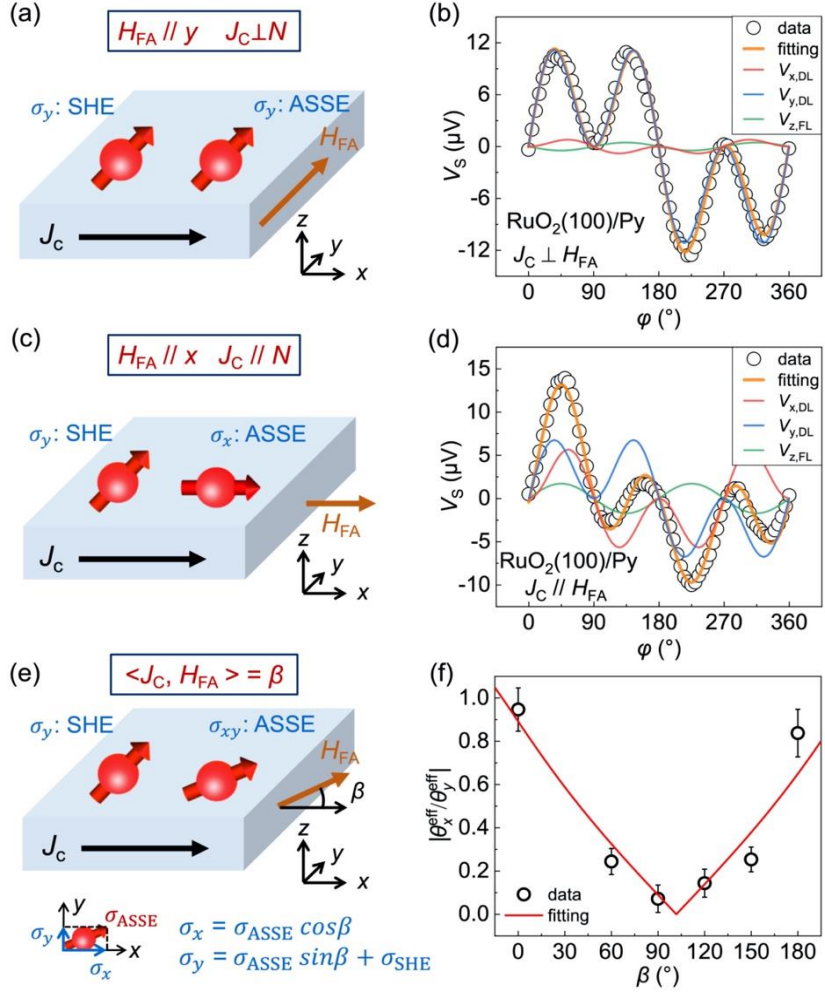

FIG. 3. (a, c, e) Schematic of spin current generation in (100)-oriented $RuO_2$ with (a) $J_C \perp H_{FA}$, (c) $J_C // H_{FA}$ or (e) $<J_C, H_{FA}> = \beta$ (The angle between $J_C$ and $H_{FA}$ is $\beta$). $J_C$ and $H_{FA}$ represent the charge current and the annealing field, respectively. (b, d) Angle-dependent $V_S$ of ST-FMR data for configurations of (b) $J_C \perp H_{FA}$ and (d) $J_C // H_{FA}$. Orange lines are fitted by the Eq. (S3). Red, blue and green lines represent angle-dependent voltage signals contributed by the damping-like torque of $\sigma_x$, $\sigma_y$ as well as the field-like torque of $\sigma_z$, respectively. $\sigma_i$ represents the $i$-axis spin polarization. (f) Ratios of $|\theta_x^{eff}/\theta_y^{eff}|$ as a function of $\beta$. $\theta_i^{eff}$ represents the spin

torque efficiency of $\sigma_i$. Red line is fitted by the expression of $|\cos\beta/(\sin\beta+C)|$. The specific positions of $J_C \perp H_{FA}$ and $J_C // H_{FA}$ for $\beta = 90°$ and $180°$, respectively, are highlighted. The error bars are from three repetitive measurements.

To further investigate the SST in $RuO_2$, magnetic field annealing was carried out in $RuO_2$(100)/Py sample to align the Néel vector along the direction of annealing field ($H_{FA}$) [25, 29]. Subsequently, we performed $\varphi$-dependent ST-FMR measurements for the scenario of charge current ($J_C$) perpendicular, parallel and at an angle $\beta$ to $H_{FA}$, the results are shown in Figure 3. For the case of $J_C \perp H_{FA}$, the angular dependence of $V_S$ can be fitted by $\sim\sin2\varphi\cos\varphi$ [Fig. 3(b)], indicating the major contribution from $y$-axis spin polarization ($\sigma_y$). This is supported by the line-shape separation result that the amplitude of $V_{y,DL}$ (blue line) is much larger than that of $V_{x,DL}$ (red line). The ratio of $|\theta_x^{\text{eff}}/\theta_y^{\text{eff}}|$ (equivalent to the $|V_{x,DL}/V_{y,DL}|$ [25]) is ~0.07 in this case. Here, $\sigma_i$ represents the $i$-axis spin polarization, $V_{i,FL(DL)}$ refers to the voltage signal contributed by the field-like (damping-like) torque of $\sigma_i$, $\theta_i^{\text{eff}}$ represents the spin torque efficiency of $\sigma_i$. The scenario differs dramatically when $J_C // H_{FA}$. Corresponding data in Fig. 3(d) are not in consistent with the angular dependence $\sim\sin2\varphi\cos\varphi$. By the fitting based on Eq. (S3), it is found that the amplitude of $V_{x,DL}$ is comparable to that of $V_{y,DL}$ ($|\theta_x^{\text{eff}}/\theta_y^{\text{eff}}|$ is ~0.84), revealing that obvious $x$-axis spin polarization ($\sigma_x$) emerges. The results above are consistent with SST that spin polarization direction is parallel to the Néel vector, as depicted by the schematics in Fig. 3(a) and 3(c). In the case of $J_C \perp H_{FA}$, both SHE and ASSE induced spins are along $y$-axis, and $\sigma_x$ is absent. While for the scenario of the $J_C // H_{FA}$, $\sigma_x$ is generated by the ASSE.

When the charge current is at an angle $\beta$ to $H_{FA}$, spin polarization generated by ASSE has both $x$-axis and $y$-axis components [Fig. 3(e)], giving rise to $\beta$-dependent ratio of $\theta_x^{\text{eff}}/\theta_y^{\text{eff}}$. As analyzed in Fig. S16, $|\theta_x^{\text{eff}}/\theta_y^{\text{eff}}|\sim|\cos\beta/(\sin\beta+C)|$, where the constant $C$ represents the ratio of $\theta_{SHE}^{eff}/\theta_{ASSE}^{eff}$ [25]. Here, $\theta_{SHE}^{eff}$ and $\theta_{ASSE}^{eff}$ represent the spin torque efficiency induced by SHE and ASSE, respectively. We plot in Fig. 3(f) the experimental ratios of $|\theta_x^{eff}/\theta_y^{eff}|$ at different $\beta$, which can be well fitted by the equation above. This result further supports the conclusion that SST exists in RuO$_2$(100) films. The minimum of the fitting line deviates slightly from 90°, which is ascribed to the fitting error and the angle error in the device fabrication. Note that $V_{z,FL}$ term with angular dependence $\sim\sin2\varphi$ (green line) appears in Fig. 3(d), which could originate from the spin precession [19–21, 30]. Angle-dependent ST-FMR measurements were also carried out in RuO$_2$(110)/Py sample with magnetic field annealing. In this case, $\sigma_x$ is negligible, which is consistent with the theoretical analysis [12, 25].

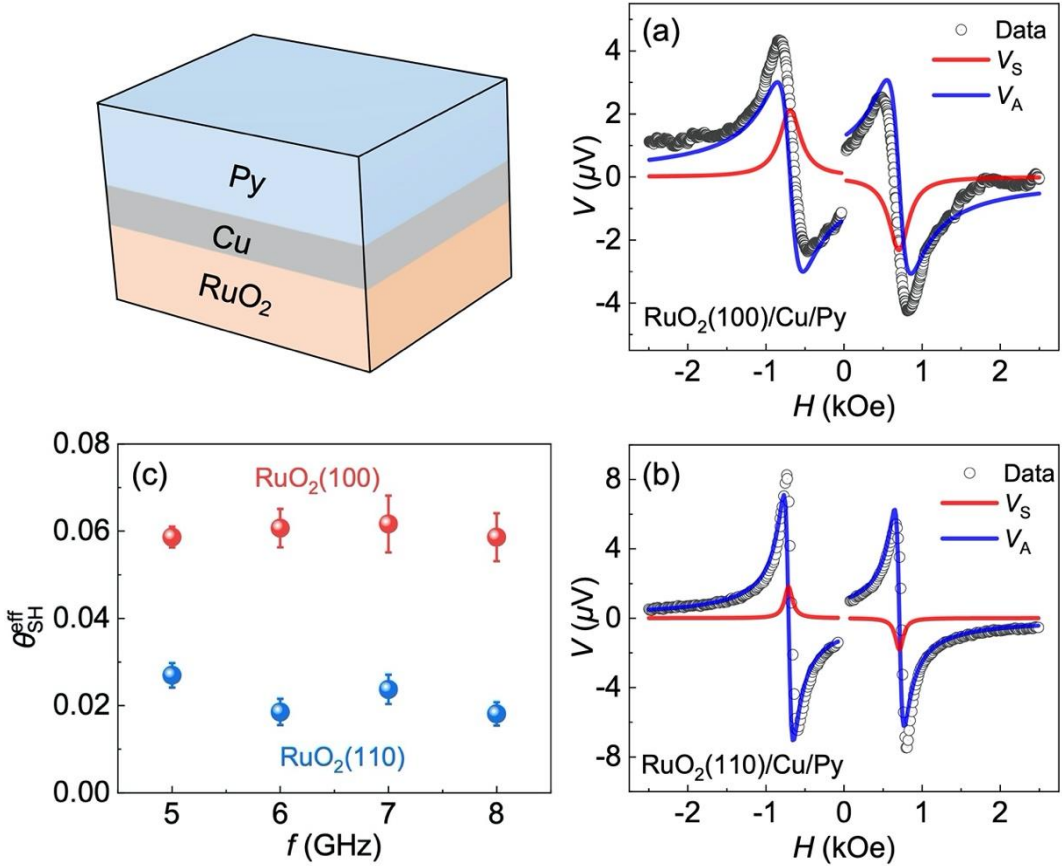

FIG. 4. ST-FMR measurement results for the samples with 2 nm thick Cu insertion. (a, b) ST-FMR spectra of (a) RuO$_2$(100)/Cu/Py and (b) RuO$_2$(110)/Cu/Py samples. Red

and blue lines are the symmetric ($V_S$) and the antisymmetric ($V_A$) components. (c) Calculated $\theta_{SH}^{eff}$ at different microwave frequencies for the RuO$_2$(100)/Cu/Py and the RuO$_2$(110)/Cu/Py samples.

Besides SST, magnetic moment-dependent spin torque behaviors can be induced by the magnetic (antiferromagnetic) spin Hall effect (MSHE or AFM-SHE), where the interfacial effect (*e.g.*, interfacial spin precession) plays a key role [19, 21, 31–34]. Differently, the origin of the $\mathcal{T}$-odd SST is ascribed to the SOC-independent magnetic exchange interaction, where the bulk contribution is dominant. To exclude the influence of interfacial effect, we show in Fig. 4 the ST-FMR data of RuO$_2$ (12 nm)/Cu (2 nm)/Py (8 nm) samples. As presented in Fig. 4(a) and 4(b), the $V_S/V_A$ ratio in (100)-oriented RuO$_2$ sample is much larger than that of (110)-oriented one, indicating the spin torque efficiency of the former is larger than that of the latter. The calculated $\theta_{SH}^{eff}$ at different frequencies are displayed in Fig. 4(c), which exhibit weak dependence on microwave frequency for both samples. $\theta_{SH}^{eff}$ of RuO$_2$(100)/Cu/Py sample is about 0.06, much larger than that of RuO$_2$(110)/Cu/Py sample ($\theta_{SH}^{eff}$~0.02), which demonstrates that the RuO$_2$ bulk dominates the generation of SST. This conclusion is supported by the angle-dependent ST-FMR measurements [25]. We point out that the slight reduction of $\theta_{SH}^{eff}$ compared to the counterpart without Cu insertion is reasonable because the additional interface and Cu spacer may increase the spin loss.

Based on the clues described above (*i.e.,* crystal orientation dependent spin torque efficiency, Néel vector dependent spin polarization direction and bulk dominant contribution), we claim that SST is observed in the RuO$_2$(100)/Py sample. Then we tend to compare the present ASSE and previous AFM-SHE [21]. Phenomenologically, ASSE-induced spin current generation in antiferromagnetic RuO$_2$ is the AFM-SHE-like behavior, because both of them show Néel vector dependent charge-spin conversion. Nevertheless, strictly speaking, ASSE in $RuO_2$ is different from AFM-SHE due to the former is dependent on nonrelativistic crystal structure (spin band) rather than the relativistic spin-orbit coupling.

Apart from the controllable spin polarization, SST in $RuO_2$ brings about ultrahigh spin torque conductivity, making it a promising spin source in spintronics. Combining the predicted ultrahigh charge-spin conversion efficiency [12] and ultralow resistivity measured in single crystal [35], the maximum $\sigma_{SH}^{eff}$ of $RuO_2$(100) is as high as $8 \times 10^5$ $\hbar/2e$ $\Omega^{-1}$ $m^{-1}$, which is larger than that of typical heavy metals and topological insulators, *e.g.*, Pt ~ $3.4 \times 10^5$, $\beta$-Ta ~ $8 \times 10^4$, $Bi_2Se_3$ ~ $2 \times 10^5$ $\hbar/2e$ $\Omega^{-1}$ $m^{-1}$ [7, 26, 36]. Such a large value is ascribed to the nonrelativistic anisotropic spin splitting effect. In addition, some other interesting physical behaviors including the crystal Hall effect [37, 38], the giant tunneling magnetoresistance [39, 40], the strain-induced superconductivity [41, 42] as well as the spin-valley locking [43] were predicted theoretically or reported experimentally in rutile $RuO_2$. All these intriguing findings make $RuO_2$ an emergent material in condensed matter physics.

In summary, we provide experimental evidences for the observation of spin splitting torque in a collinear antiferromagnet $RuO_2$, where three typical features of SST were observed: (i) spin torque efficiency of $RuO_2$(100) film is much larger than that of $RuO_2$(110) film, because of the crystal orientation-dependent spin current flowing along [100]-axis of $RuO_2$; (ii) the direction of spin polarization generated by $RuO_2$(100) film is dependent on (parallel to) the Néel vector of $RuO_2$; (iii) spin torque efficiencies measured in $RuO_2$(100)/Py samples with and without Cu insertion are comparable, and both are larger than that of $RuO_2$(110) samples, demonstrating the bulk contribution is dominant. Spin splitting torque in $RuO_2$ brings about controllable spin polarization combined with ultrahigh spin torque conductivity, which makes $RuO_2$ a promising spin source in spintronics.

**Note Added**——After finishing the experimental work and during the manuscript preparation, we are aware of a relevant work that demonstrated (101)-oriented $RuO_2$ can generate out-of-plane spin polarization [44]. In this work, we mainly focus on the spin splitting torque generated in (100)-oriented $RuO_2$.

This work is supported by the National Key R&D Program of China (Grant No. 2021YFB3601301), the National Natural Science Foundation of China (Grant No. 51871130) and and the Natural Science Foundation of Beijing, China (Grant No. JQ20010). C.S. acknowledges the support of Beijing Innovation Center for Future Chip (ICFC), Tsinghua University.

**Reference**

*These authors contributed equally to this work.

†Corresponding author:

fanxiaolong@lzu.edu.cn;

‡Corresponding author:

songcheng@mail.tsinghua.edu.cn

RF current and Oersted field, frequency-dependent FMR measurements, data of angle-dependent ST-FMR measurements, key values of ST-FMR data in Fig. 2 and Fig. 4, detailed analysis of SST in $RuO_2(100)$/Py sample and the comparison of ASSE and AFM-SHE, which includes Refs. [12, 17, 18, 21, 26, 29, 44–50].